\documentclass[lettersize,journal]{IEEEtran}
\usepackage{amsmath,amsfonts}
\usepackage{algorithmic}
\usepackage{array}
\usepackage{textcomp}
\usepackage{stfloats}
\usepackage{url}
\usepackage{verbatim}
\usepackage{graphicx}
\usepackage{subcaption}

\def\BibTeX{{\rm B\kern-.05em{\sc i\kern-.025em b}\kern-.08em
    T\kern-.1667em\lower.7ex\hbox{E}\kern-.125emX}}
\usepackage{balance}
\begin{document}
\title{Physics-Informed Deep Neural Network Design of Reactively Loaded Metasurfaces}
\author{Malik Almunif, \IEEEmembership{Graduate Student Member, IEEE}, John Le, \IEEEmembership{Graduate Student Member, IEEE},  and Anthony Grbic, \IEEEmembership{Fellow, IEEE}
\thanks{The authors are with the Electrical and Computer Engineering Department, University of Michigan, Ann Arbor, MI 48109 USA (e-mail: agrbic@umich.edu).}}

\maketitle
\begin{abstract}
A tandem deep neural network approach is presented for the inverse design of reactively loaded metasurfaces with prescribed far-field radiation characteristics. The proposed approach integrates a deep neural network (DNN) with a physics-based microwave network forward solver. The DNN maps target far-field patterns to distributions of reactive loads across the metasurface unit cells. The predicted distribution of reactive loads is evaluated by the forward solver to compute the resulting radiation pattern and guide the learning process through a cosine-similarity loss function. The forward solver enables a fast evaluation of the metasurface's electromagnetic response, significantly reducing the computational cost required for training. The proposed approach is applied to a metasurface with aperture-coupled unit cells loaded with reactances. Several design examples are presented to demonstrate the accurate synthesis of shaped and steered radiation patterns. Full-wave electromagnetic simulations are performed to validate the accuracy of the designed beamforming metasurfaces.
\end{abstract}

\begin{IEEEkeywords}
Antenna, beamforming, machine learning, metasurface, network theory, neural network.
\end{IEEEkeywords}

\section{Introduction}

\IEEEPARstart{I}{nverse} design is an important tool in the synthesis of metasurfaces, allowing arbitrary control of electromagnetic fields \cite{n1}. Metasurfaces have been extensively employed in a wide range of applications, including polarization control \cite{C1}, beamforming \cite{a1}, \cite{a2}, \cite{N1}, \cite{a4}, \cite{a5}, cloaking \cite{a6}, and many others.\par
Metasurface synthesis techniques typically rely on iterative optimization employing forward solvers based on various techniques including the Method of Moments (MoM) \cite{a7}, \cite{a8}, \cite{a9}, \cite{N3}, finite element method \cite{ag11}, and even microwave network theory \cite{a3}, \cite{a10}, \cite{N2}. In synthesis, a target far-field radiation pattern or near-field profile is specified, and the spatial variation of parameters such as dimensions or impedances are optimized to achieve the target response. While effective for small apertures or fixed designs, these techniques become computationally prohibitive for electrically-large metasurfaces or scenarios requiring real-time beamforming. \par
To overcome the aforementioned challenges, machine learning has emerged as a  promising tool for accelerating metasurface inverse design \cite{an11}. In particular, deep neural networks (DNNs) can learn complex nonlinear mappings between target field responses and device parameters, enabling rapid synthesis without repeated full-wave simulations after training is complete. Prior work, such as \cite{a11}, proposed the use of rotatable dielectric elements to achieve real-time beamforming. However, the required training data is generated through full-wave simulations, which are computationally expensive. In \cite{a12}, transmissive metasurfaces capable of producing specified far-field masks were designed using a DNN-based approach. This method first reconstructs an equivalent source array that reproduces the target pattern and then attempts to realize it with a physical metasurface. This procedure introduces certain limitations. In particular, the local power conservation constraint must be enforced on the output source array to avoid active or lossy unit cells. Additionally, the realization of the output source array uses a local periodicity approximation which only approximates mutual coupling, leading to degraded performance.  \par

In this letter, a tandem deep neural network approach is proposed for the inverse design of reactively loaded metasurfaces. The approach integrates a convolutional neural network with a physics-based forward solver formulated using microwave network theory. A single full-wave simulation is required prior to training in order to extract the impedance matrix and active element patterns of the metasurface. These quantities are subsequently used by the forward solver to rapidly predict the far-field pattern for an arbitrary distribution of reactive loads during training. Eliminating the need for repeated full-wave simulations to generate training data significantly reduces its overall computational cost. Consequently, the inverse network can be trained using randomly generated radiation patterns, with the learning process guided by the forward solver. Several design examples are presented to demonstrate the accuracy of the trained inverse network in the  synthesizing of metasurfaces.  \par

\section{Review of the Reactively Loaded Metasurface Design}
\noindent The metasurface considered in this letter consists of a grid of aperture-coupled unit cells \cite{a13}, as illustrated in Fig. 1. Each unit cell comprises three layers. The top layer consists of metallic patches, the middle layer is a ground plane perforated with slots to enable coupling between the top and bottom layers, and the bottom layer consists of microstrip lines terminated  with lumped ports. The lumped ports allow the entire metasurface to be characterized in full-wave simulation by its impedance matrix and the active element patterns of the individual unit cells \cite{a3}. Once the metasurface is characterized, one unit cell is directly-fed while all remaining cells are terminated with reactive loads. This configuration enables reactive loads to operate as parasitic radiators, whose induced currents shape the overall radiation pattern. \par

\begin{figure}[!t]\centerline{\includegraphics[]{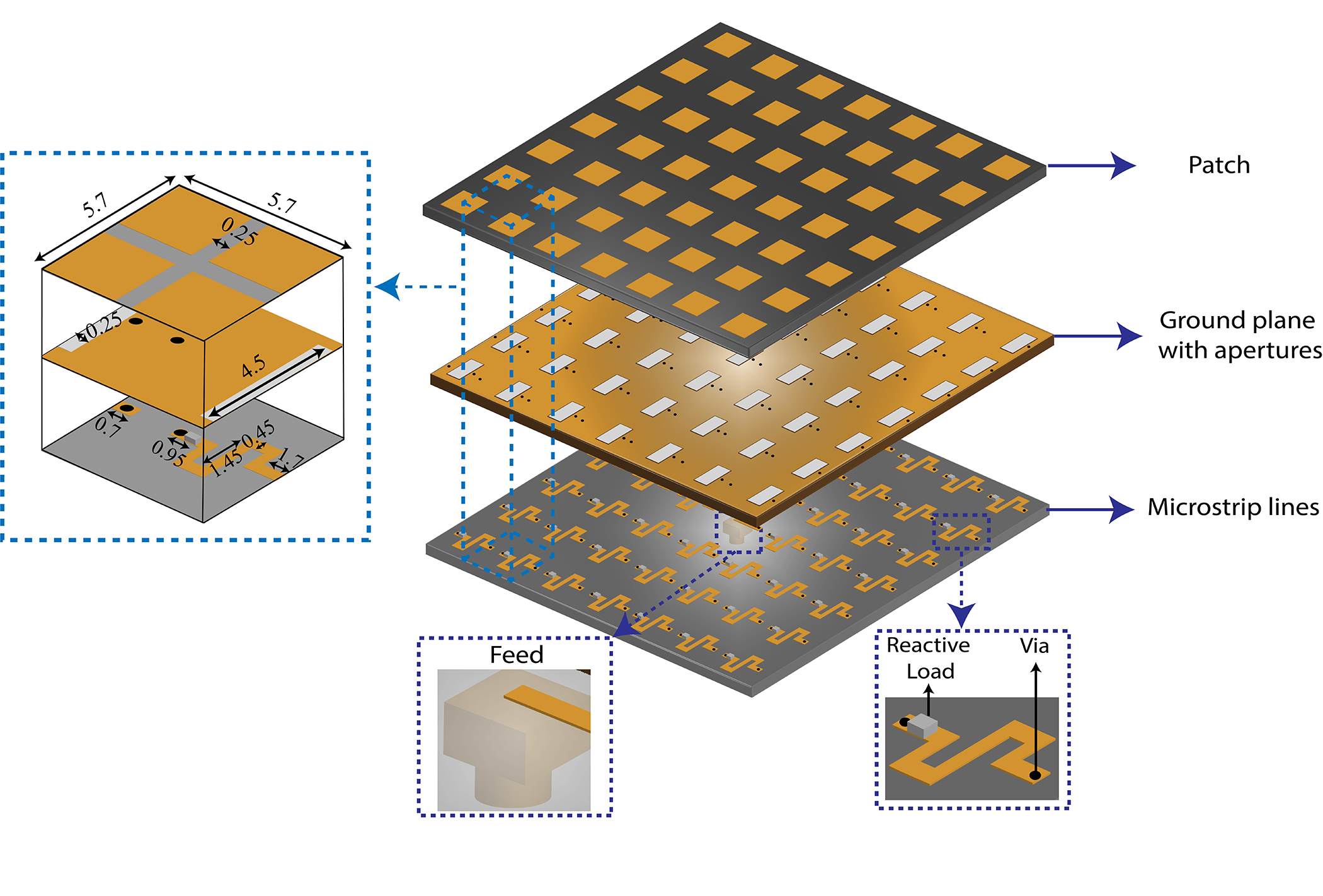}}
        \caption{Aperture-coupled metasurface composed of a two-dimensional grid of unit cells. All cells are terminated with reactive loads, except for one centrally located unit cell that is directly-fed to excite the metasurface. All dimensions are in mm.}
\label{Fig1}
\end{figure}

 The total radiated far-field from the metasurface, $\mathbf{E_{tot}(\theta,\phi)}$, is a superposition of the active element patterns of all unit cells weighted by their port currents. It can be expressed as, 
\begin{equation}
    \mathbf{E_{tot}(\theta,\phi)}=\displaystyle \sum_{m=1}^{M} I_m \mathbf{E_m(\theta,\phi)}
    \label{Edes}
\end{equation}
where $I_m$ is the current at the $m$th port, and $\mathbf{E_m(\theta,\phi)}$ represents the active element pattern of the $m$th unit cell. \par
The port currents corresponding to a set of reactive loads can be determined using microwave network theory. The microwave network representation of the metasurface is given by \cite{a3}
\begin{equation}
\begin{bmatrix}
\Bar{0}\\
\Bar{V}^E\\
\end{bmatrix}
=
\begin{bmatrix}
\Bar{\Bar{Z}}^{LL}+j\Bar{\Bar{X}}_L&\Bar{\Bar{Z}}^{LE}\\
\Bar{\Bar{Z}}^{EL}&\Bar{\Bar{Z}}^{EE}
\end{bmatrix}
\\
    \begin{bmatrix}
\Bar{I}^L\\
\Bar{I}^E
\end{bmatrix}
\label{eq1}
\end{equation}
where the superscripts $L$ and $E$ denote the terminated (loaded) and excited unit cells, respectively. The $Z$-matrices are computed in advance using a full-wave solver. The matrix $\Bar{\Bar{X}}_L$ is diagonal with entries equal to the reactive loads. Solving the first row of (\ref{eq1}) yields the current induced in the reactive loads, 
\begin{equation}
\Bar{I}^L=-[j\Bar{\Bar{X}}_L+\Bar{\Bar{Z}}^{LL}]^{-1}\Bar{\Bar{Z}}^{LE}\Bar{I}^E
\label{IL}
\end{equation}
This relation links the distribution of reactive loads to the total far-field radiated by the metasurface via the induced currents and active element patterns. Consequently, by appropriately selecting the reactive loads, a wide range of radiation patterns can be synthesized. Additionally, the input impedance at the directly-fed unit cell can be derived from the second row of (\ref{eq1}) \cite{a3},

\begin{equation}
    \Bar{\Bar{Z}}_{in}=\Bar{\Bar{Z}}^{EL}[-j\Bar{\Bar{X}}_L-\Bar{\Bar{Z}}^{LL}]^{-1}\Bar{\Bar{Z}}^{LE}+\Bar{\Bar{Z}}^{EE}
    \label{S11}
\end{equation}
The input impedance matrix can be converted to an S-matrix using standard microwave network conversions \cite{apozar}. This microwave network formulation is integrated inside the proposed tandem deep neural network to enable the inverse network to learn how to synthesize target far-field patterns.

\section{Deep Neural Network}
\begin{figure*}[!t]\centerline{\includegraphics[scale=0.85]{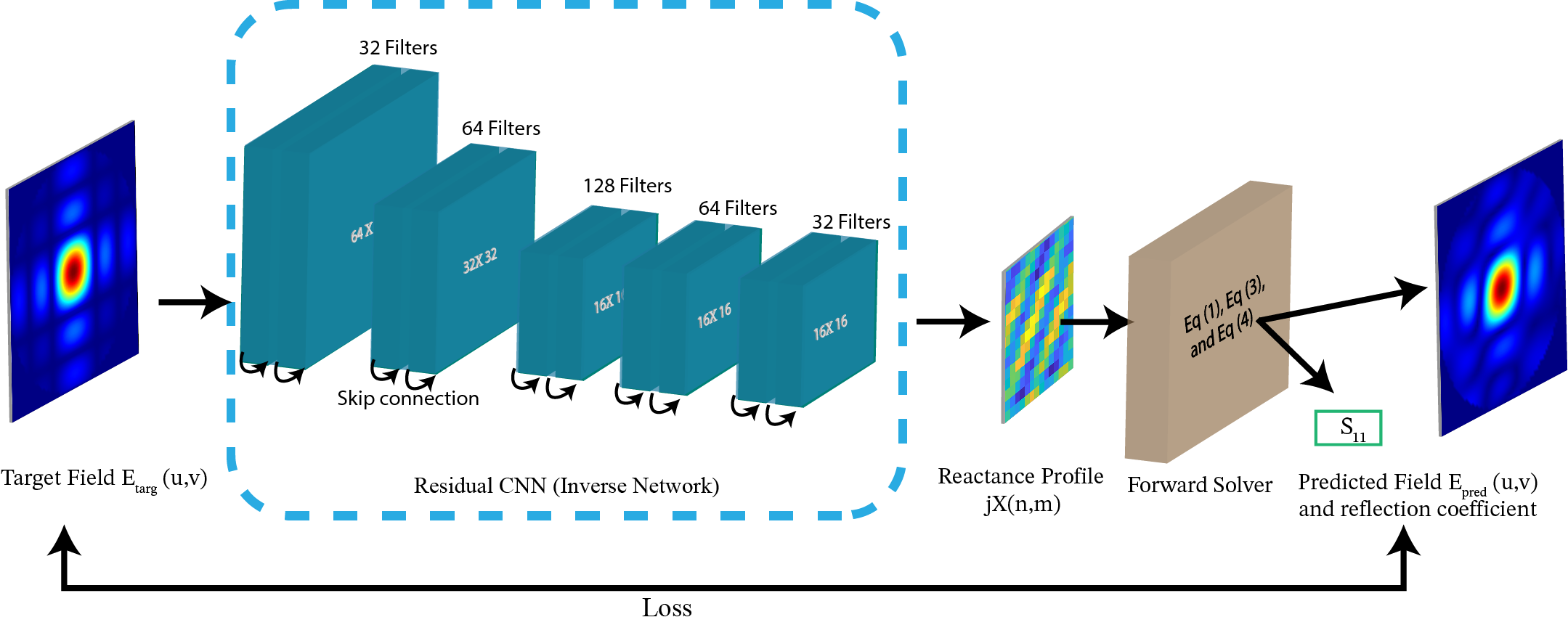}}
\caption{ Proposed tandem deep neural network architecture. The inverse network maps a target far-field pattern in the $u$-$v$ plane to a distribution of reactive loads connected to the ports of the metasurface. The predicted distribution of reactive loads is then passed to a microwave network forward solver to compute the resulting radiated far-field and reflection coefficient. }
\label{Fig3}
\end{figure*}

\noindent In this section, a tandem deep neural network is introduced for the automated design of reactively loaded beamforming metasurfaces. The proposed approach, illustrated in Fig. \ref{Fig3}, combines a data-driven inverse network with a physics-based microwave network forward solver. The building blocks of the tandem network are described in the following subsections.

\subsection{Inverse Network}
\noindent

\noindent The inverse network is implemented as a deep convolutional neural network (CNN) with residual blocks that  maps a target far-field pattern to a corresponding distribution of reactive loads. The input to the CNN is a far-field pattern sampled in the $u$-$v$ plane and represented as a two-dimensional image of size $64 \times 64$. The output is a distribution of reactive loads $jX(n,m)$ with dimensions $15 \times 15$, where $n$ and $m$ are indices that identify the unit cell of the metasurface. \par

The CNN consists of stacked convolutional layers that extract spatial features from the input far-field pattern. These layers employ small ($3\times 3$) convolutional kernels (filters) to capture key radiation characteristics such as main beam direction, sidelobe levels, and null locations. As the network depth increases, the number of filters expands, enabling the network to learn broader features. To improve training stability and enable deeper feature extraction, the network incorporates residual blocks with skip connections \cite{a14}. These skip connections are used to prevent vanishing gradients \cite{a14}.\par 
The spatial resolution of the feature maps is progressively reduced through strided convolutions, resulting in an intermediate output of size $16\times16$. This output is then centrally cropped to match the metasurface grid size. A LeakyReLU activation function is used for all the intermediate layers, while a sigmoid activation function is applied at the final layer. The normalized output of the sigmoid is subsequently scaled to the desired reactance range $[-150, -1]\Omega$, ensuring that all predicted loads are capacitive. This constrains the range of reactive loads to those achievable by varactors (tunable reactances), making the designs amenable to a reconfigurable implementation. The resulting predicted distribution of reactive loads is then provided as input to the forward solver.

\subsection{Forward Solver}
\noindent The forward solver takes the predicted distribution of reactive loads as input and computes the corresponding far-field pattern and the reflection coefficient using the microwave network formulation described in Section II \cite{a3}. Given a distribution of reactive loads, the solver first computes the induced port currents using \eqref{IL} and then evaluates the resulting far-field pattern using \eqref{Edes}. In addition, the reflection coefficient at the directly-fed unit cell is calculated from the corresponding input impedance, as expressed in (\ref{S11}). The forward solver is implemented using differentiable TensorFlow operations, allowing gradients to backpropagate through the microwave network equations during training. This enables end-to-end training of the physics-informed tandem networks. \par
The microwave network-based solver requires approximately $1.2$ seconds for a single evaluation of the far-field pattern and reflection coefficient, compared to $12.2$ hours for a full-wave electromagnetic simulation in Ansys HFSS. This reduction in computational cost enables rapid forward evaluation during network training.

\subsection{Training Data}
\noindent
The proposed tandem network eliminates the need for paired datasets of reactive load distributions and far-field patterns during training, since the forward solver evaluates the electromagnetic response using a microwave network formulation. Instead, only random far-field patterns are required. To ensure realistic training data, random far-field patterns are generated using classical array theory \cite{a15} by assuming an array with the same aperture size as the metasurface. These patterns are obtained by assigning different amplitude and phase excitations to the array elements and computing the corresponding array factor. The resulting dataset spans a wide range of steering angles, beamwidths, and sidelobe levels, enabling the inverse network to generalize across diverse beamforming objectives. To further increase the size and diversity of the training dataset, data augmentation is employed by randomly mixing the generated radiation patterns. In total, $50000$ radiation patterns are generated.  The dataset is randomly divided into training and testing sets using an $80/20$ split, resulting in $40000$ patterns for training and $10000$ patterns for testing. \par 

During training, batches of target far-field patterns are fed into the inverse network to predict the corresponding distribution of reactive loads. These reactive loads are evaluated by the forward solver to compute the resulting far-field patterns and reflection coefficients. The predicted and target responses are then compared using the following loss function, 
\begin{align}
\mathrm{Loss} =&
\Bigg[1 -
\frac{1}{B}
\sum_{b=1}^{B}
\frac{
\left\langle 
E_{\mathrm{targ}}^{(b)},\;
E_{\mathrm{pred}}^{(b)}
\right\rangle
}{
\left\|E_{\mathrm{targ}}^{(b)}\right\|_2\;
\left\|E_{\mathrm{pred}}^{(b)}\right\|_2
}\Bigg] \nonumber \\
&+
\frac{1}{B}
\sum_{b=1}^{B}
\max\!\left(0,\;
S_{11}^{\mathrm{pred} (b)} - S_{11}^{\mathrm{threshold}}
\right)
\label{eq:beam_loss}
\end{align}
where $B$ denotes the batch size, $\langle \cdot \rangle$ represents the inner product between the target and predicted far-field patterns, and  $\|\cdot\|_2$ denotes the $\ell_2$ norm. The first term corresponds to a cosine-similarity loss between the target and predicted far-field patterns, emphasizing agreement in far-field pattern shape. The second term penalizes reflection at the directly-fed unit cell when the predicted reflection coefficient exceeds a prespecified threshold of $S_{11}^{\mathrm{threshold}}=0.3$. The proposed neural network is  trained using TensorFlow with the Keras API and the AdamW optimizer.

\section{Design Examples}
\noindent
In this section, the trained neural network is used to generate several radiation patterns using the beamforming metasurface in order to demonstrate the capabilities of the proposed approach. Three target radiation patterns are considered, each with a distinct beamforming objective.\par
The first target far-field pattern corresponds to that of a uniformly excited array with the main beam steered towards $(\theta_0,\phi_0)=(-30^\circ, 45^\circ)$. The target far-field pattern is computed using the array factor as follows, 
\begin{multline} 
E_{\mathrm{targ}}(u,v)
=\\\displaystyle \sum_{n=-(N_x-1)/2}^{(N_x-1)/2} \displaystyle \sum_{m=-(N_y-1)/2}^{(N_y-1)/2}
e^{
j k_0 d\left(
n (u - u_0) + m (v - v_0)
\right)
}
\label{eq:uniform_AF}
\end{multline}
where $d$ is the unit cell spacing, $N_x$ and $N_y$ are the total number of unit cells along $x$ and $y$, $u_0=\sin\theta_0 \cos \phi_0$, and $v_0 =\sin\theta_0\sin\phi_0$.\par
The second example aims to produce a flat-top far-field pattern oriented along the diagonal of the $u$-$v$ plane, representing a more challenging beam shaping objective. The target pattern is defined by a piecewise function,

\begin{equation}
E_{\mathrm{targ}}(u,v)
=
\begin{cases}
1, & |u'|\leq 0.7 \ \text{and}\ |v'|\leq 0.2, \\
0, & \text{otherwise},
\end{cases}
\label{eq:flattop_uv}
\end{equation}
where $u'$ and $v'$ are rotated coordinates given by, 
\begin{equation}
\begin{bmatrix}
u' \\
v'
\end{bmatrix}
=
\begin{bmatrix}
\cos 45^\circ & \sin 45^\circ \\
-\sin 45^\circ & \cos 45^\circ
\end{bmatrix}
\begin{bmatrix}
u \\
v
\end{bmatrix}
\end{equation}

In the third example, the target far-field pattern is a multibeam pattern, with beams pointing towards $(\theta_0,\phi_0)=(-30^\circ,90^\circ )$ and $(\theta_0,\phi_0)=(-30^\circ,-45^\circ )$. The target pattern is constructed as a superposition of two uniformly steered array factors defined in (\ref{eq:uniform_AF}).\par
\begin{figure*}[!t]\centerline{\includegraphics[scale=0.83]{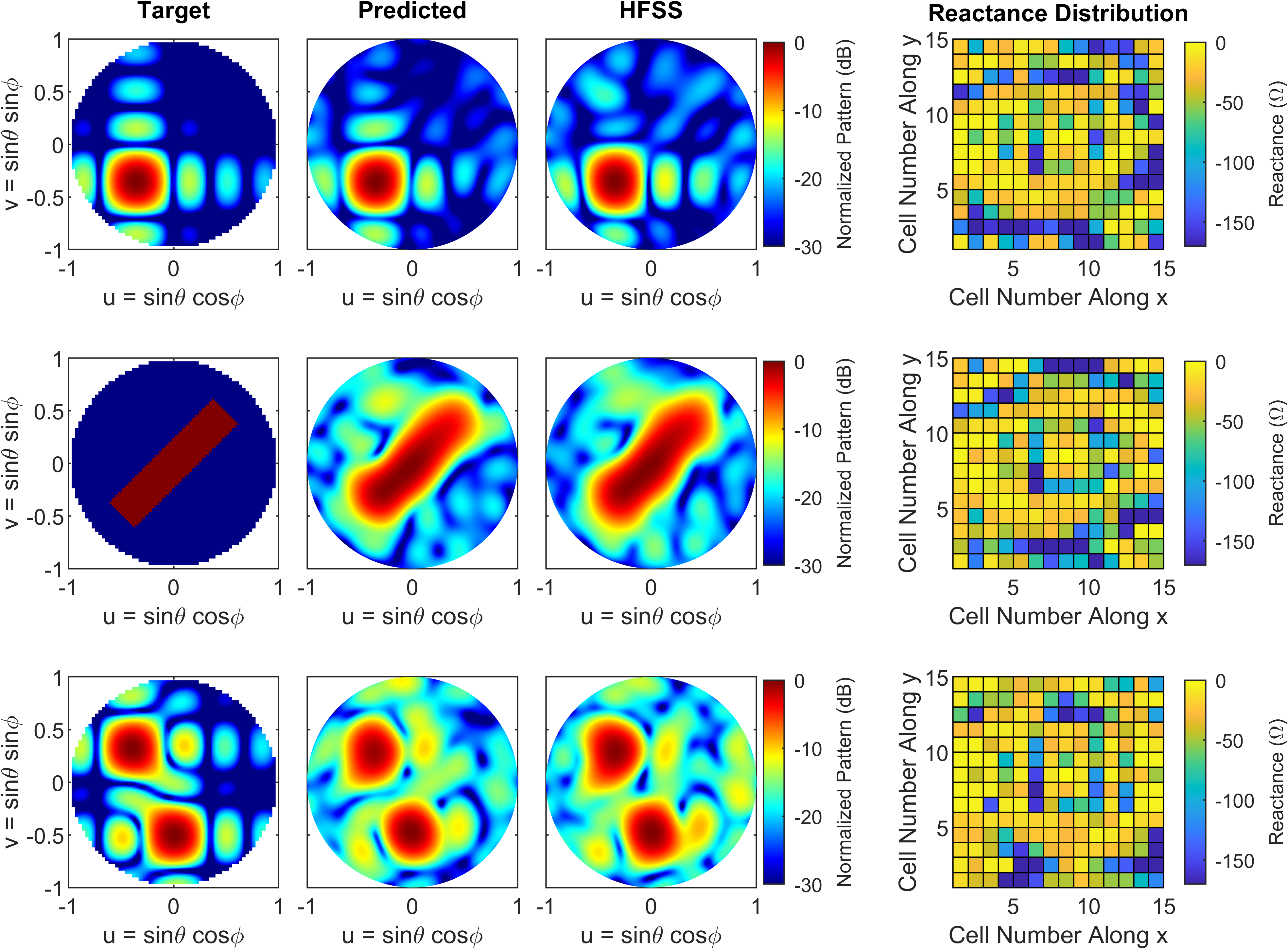}}
\caption{Three example metasurfaces designed using the proposed inverse network. The first example corresponds to a uniformly excited array steered off the principle planes. The second example radiates a flat-top far-field pattern oriented along the diagonal of the $u$-$v$ plane. The third example radiates a multibeam far-field pattern. }
\label{Fig4}
\end{figure*}

 
The three design examples are shown in Fig. \ref{Fig4}. For each case, the target far-field pattern is provided as input to the inverse network, which predicts the corresponding distribution of reactive loads across the metasurface unit cells. The predicted reactive loads are then evaluated using the microwave network forward solver to compute the resulting, predicted far-field pattern. To further validate the metasurface designs, the predicted reactive loads are attached to the metasurface using lumped capacitances in the full-wave solver Ansys HFSS.  As observed in Fig. \ref{Fig4}, the synthesized far-field patterns exhibit close agreement with the corresponding target patterns in all cases, demonstrating the precision of the proposed inverse network. The slight discrepancy observed at  the edges of the flat-top pattern is attributed to the finite aperture size of the metasurface. In addition, the calculated reflection coefficients at the directly-fed unit cell are summarized in Table I for all design examples. In all cases, the directly-fed unit cell is impedance matched to a $50$ $\Omega$ source. \par

\begin{table}[!t]
\caption{Reflection coefficient at the directly-fed unit cell.}
\centering
\begin{tabular}{|c|c|c|}
\hline
Example Number & Predicted $S_{11}$ (dB) &Simulated $S_{11}$  (dB)\\
\hline
Design 1 & $-10.71$ & $-10.3$\\
\hline
Design 2 & $-11.17$ & $-11.4$\\
\hline
Design 3 & $-9.56$ & $-9.54$\\
\hline
\end{tabular}
\end{table}

\section{Conclusion}
\noindent This paper presents an inverse design approach for reactively loaded beamforming metasurfaces using a physics-informed neural network. The proposed approach employs a tandem architecture consisting of a convolutional neural network followed by a microwave network-based forward solver. The convolutional neural network predicts the  distribution of reactive loads from a target far-field pattern. The predicted reactive loads are subsequently evaluated by the forward solver to compute the resulting far-field pattern. The network is trained by comparing the predicted and target far-field patterns using a cosine-similarity loss function. Several design examples are presented to demonstrate the effectiveness of the proposed inverse network in synthesizing beamforming metasurfaces with different target radiation patterns.

\section*{ACKNOWLEDGEMENT}
\noindent The authors thank King Abdulaziz City for Science and Technology (KACST) for providing M. Almunif with a Ph.D. scholarship and DEVCOM Army Research Laboratory for providing J. Le with a DEVCOM ARL Summer Student Experience.

\end{document}